\documentclass[fleqn]{mnras}

\usepackage{mathptmx}

\usepackage[T1]{fontenc}
\usepackage{ae,aecompl}

\usepackage{graphicx}    
\usepackage{amsmath}    
\usepackage{bm}
\usepackage{amssymb}    

\newcommand{\be}{\begin{equation}}
\newcommand{\ee}{\end{equation}}
\newcommand{\varv}{\mbox{\em v}}

\title[Sub-structure formation in starless cores]{Sub-structure formation in starless cores}

\author[C. Toci et al.]{
C.~Toci,$^{1,2}$\thanks{E-mail: claudia@arcetri.astro.it}
D.~Galli,$^{2,1}$
A.~Verdini$^{1,2}$
L.~Del Zanna$^{1,2,3}$
S.~Landi$^{1,2}$
\\
$^1$Universit\`a degli Studi di Firenze, Via G. Sansone 1, I-50019, Sesto Fiorentino, Italy\\
$^2$INAF - Osservatorio Astrofisico di Arcetri, Largo E. Fermi 5, I-50125, Firenze, Italy\\
$^3$INFN - Sezione di Firenze, Via G. Sansone 1, I-50019, Sesto Fiorentino, Italy\\
}

\date{Accepted XXX. Received YYY; in original form ZZZ}

\pubyear{2017}

\begin{document}
\label{firstpage}
\pagerange{\pageref{firstpage}--\pageref{lastpage}}
\maketitle

\begin{abstract}
Motivated by recent observational searches of sub-structure in starless molecular cloud
cores, we investigate the evolution of density perturbations on
scales smaller than the Jeans length embedded in contracting isothermal
clouds, adopting the same formalism developed for the expanding
Universe and the solar wind. We find that initially small amplitude,
Jeans-stable perturbations (propagating as sound waves in the
absence of a magnetic field), 
are amplified adiabatically during the contraction,
approximately conserving the wave action density, until they either
become nonlinear and steepen into shocks at a time $t_{\rm nl}$,
 or become gravitationally unstable when the
Jeans length decreases below the scale of the perturbations at a
time $t_{\rm gr}$. We evaluate analytically the time $t_{\rm nl}$
at which the perturbations enter the non-linear stage using a
Burgers' equation approach, and we verify numerically that this
time marks the beginning of the phase of rapid dissipation of the
kinetic energy of the perturbations. We then show that for typical
values of the rms Mach number in molecular cloud cores, $t_{\rm
nl}$ is smaller than $t_{\rm gr}$, and therefore density perturbations
likely dissipate before becoming gravitational unstable. Solenoidal modes 
grow at a faster rate than compressible modes, and may eventually
promote fragmentation through the formation of vortical structures.
 \end{abstract}

\begin{keywords}
hydrodynamics -- ISM: clouds -- ISM: kinematics and dynamics
\end{keywords}

\section{Introduction}

The growth of small-scale density perturbations during the collapse
of an interstellar cloud has been the subject of many studies aimed
at understanding the formation of stellar clusters and 
multiple stellar system by the process of fragmentation, namely
the breakup of a large cloud into clumps, cores and stellar clusters.
While early studies (Hoyle~1953; Hunter~1962, 1964; Mestel~1965a,b) 
focused on the growth of initially Jeans-unstable perturbations,
Tohline~(1980) introduced the concept of ``delayed
fragmentation'', as the onset of gravitational instability can occur only 
after the mean density of the cloud (assumed to be in free-fall collapse) 
has grown significantly, and 
the instantaneous value of the Jeans length has become smaller
than the scale of the perturbed region. Tohline~(1980) argued
that in this case the time needed for a significant amplification of the 
perturbations would be larger than the time the parent cloud has left in 
its evolution, and pointed out the role of the converging motion of the gas,
in alternative to self-gravity, as a means to amplify 
small-scale density perturbations which are initially Jeans-stable.
The fate of these initially
Jeans-stable perturbations was investigated numerically in collapse 
simulations by Rozyczka~(1983),
who found them subject to efficient damping well before reaching the
threshold for gravitational instability, and therefore unable to form well-defined fragments.
Furthermore, in his simulations Rozyczka~(1983) observed the formation 
of local concentrations of angular momentum (possibly as the result
of tidal interactions) even in the collapse of initially non-rotating clouds.
Whether these local enhancements of vorticity would turn themselves into  
turbulent eddies, as originally suggested by Layzer~(1963), or eventually evolve
to rotationally supported disk-like structures, as suggested e.g. by the simulations
by Goodwin et al.~(2004), has remained unclear.                     

Recently, an original approach to the problem of the growth of density perturbations
embedded in a collapsing cloud was adopted by Toal\`a et al.~(2015), who
formulated the problem in the framework of an {\em inverse Hubble
flow}, i.e. the gravitational collapse of a spherical pressureless
cloud, exploiting the analytical tools developed in cosmology
for the expanding Universe\footnote{In cosmology, this solution corresponds to the
Lemaitre-Tolman elliptic ($k=+1$) solution of Einstein's equations for a
``dust'' Universe}. Toal\`a et al.~(2015) found that the gravitational
instability occurs faster in inverse Hubble flows than in a static
cloud, and postulated that growing, unstable perturbations on scales
larger than the Jeans length $\lambda_{\rm J}$ collapse when they
reach the non-linear stage, i.e. at a time $t_{\rm nl}$ identified
by the condition $\delta\rho/\rho \approx 1$. In inverse
Hubble flows, $t_{\rm nl}$ is always smaller than the free-fall
time $t_{\rm ff}$ of the cloud, and approaches the latter in the
limit of initially small amplitude of the perturbations.

While Toal\`a et al.~(2015) studied the evolution of Jeans-unstable 
perturbations in a cloud containing
a large number of Jeans masses, in this work we consider molecular
cloud cores containing only a few Jeans masses, and focus on
the evolution of Jeans-stable density perturbations, that initially oscillate
as a collection of sound waves of  small amplitude. This is justified 
by the observational result that 
non-thermal motions in dense cores are generally subsonic (Myers et al.~1983).
In the case of starless cores, it has been suggested that subsonic motions 
(``turbulence'') may be
forming the seeds of multiple sites of star formation  (Fisher~2004,
Goodwin et al.~2004, 2007), but scarce evidence has been found yet for
substructure formation in these objects (Schnee
et al. 2012, Dunham et al.~2016).
For example, Schnee et al.~(2012) examined four ``super-Jeans'' starless 
cores of masses 3--8 $M_\odot$ finding upper limits on the masses of 
any embedded fragment down to a few thousandth of solar mass. 
Only recently some evidence of compact substructure has been found in one starless core
in Ophiuchus (Kirk et al.~2017) and in the Orion Molecular Cloud 1 South (Palau 
et al.~2017). In any case, the processes determining 
the formation of multiple protostellar seeds
at these scales remain unclear.

To address the problem of sub-structure formation in cloud cores,
in this paper we investigate the evolution of density and velocity
perturbations embedded in a contracting medium (the ``background'', or 
``parent cloud'')
undergoing isotropic or anisotropic collapse. First, we formulate
a theoretical framework to study contracting flows based on methods
developed in cosmology for the expanding Universe (see, e.g., Peebles~1980)
and in the study of the expanding solar wind (Grappin et
al.~1993, Grappin \& Velli~1996, Tenerani \& Velli~2013, Dong et al.~2014, 
Verdini \& Grappin~2015).  In this framework, we study analytically
the linear and nonlinear evolution of perturbations. We consider two
simple cases of contracting backgrounds: a spherical cloud undergoing
homologous pressureless collapse, and the spherical accretion flow 
on a point mass. 
 
Related problems are the evolution of oscillations in contracting
cores (Broderick \& Keto~2010, Keto \& Caselli~2010), and the
behaviour of hydrodynamical turbulence during the contraction of a
cloud (Robertson \& Goldreich~2012, Davidovits \& Fish~2017).  The
latter studies, in particular, demonstrates that the root mean square (rms) 
turbulent velocity increases 
during contraction, as long as the eddy turnover time is shorter
than the contraction time (a process termed ``adiabatic heating'' of turbulence).

The paper is organised as follows: in Sect.~2 we formulate the
equations of hydrodynamics in comoving coordinates for application
to contracting interstellar clouds; in Sect.~3 we study the evolution
of linear perturbations in the case of isotropic and anisotropic
contraction; in Sect.~4 
we apply the formalism of inverse Hubble flows to the
homologous collapse of a uniform-density cloud; in Sect.~5 we compare our 
analytical results to a hydrodynamical simulation in a contracting box. 
Finally, in Sect.~6 we summarise our
conclusions.

\section{Hydrodynamics in comoving coordinates}

To investigate the dynamics of contracting clouds, we
employ an approach similar to the one commonly used in cosmological
studies, where a background medium evolves with some specified laws
(the Friedmann equations in that case) and the growth of primordial
perturbations is followed in a local coordinate system 
comoving with the background flow. We generalise this approach by
relaxing the assumption of an isotropic expansion/contraction. Given the
non-uniformity of the interstellar medium, characterised by a filamentary 
structure and the presence of accretion flows, fluid motions are not
expected to be isotropic. For example, a fluid element accreting onto a
mass point is accelerating
and stretching in the direction of the flow, while contracting in
the transverse directions\footnote{In the case of the solar wind,
expanding at different rates in the
radial and transverse directions, a similar formalism is 
called ``expanding box model'' (Grappin et al.~1993, Grappin \&
Velli~1996).}.

Consider a small Cartesian line element 
$\delta {\bf x} = (\delta x, \delta y, \delta z)$ advected by the flow, evolving as 
\be
\delta {\bf x}(t)= {\mathbb S}(t) \cdot \delta {\bf x}_0,
\ee
where ${\mathbb S}(t)={\rm diag}[a(t), b(t), c(t)]$ is the scale factor 
normalized such that ${\mathbb S}(t_0)={\mathbb I}$ and $\delta {\bf x}(t_0)=\delta {\bf x}_0$. 
All hydrodynamical quantities can be written as the sum of background and
local components (hereafter ``perturbations"), as
\be
{\bf u}={\bf u}_{\rm b}+{\bf u}_1, \quad \rho=\rho_{\rm b}+\rho_1,
\label{expans}
\ee
where ${\bf u}({\bf x},t)$ is the velocity field, $\rho({\bf x},t)$
is the gas density, and ${\bf x} = (x,y,z)$.  In the following we
assume an isothermal equation of state 
$p=c_s^2\rho$, where $p$ is the gas pressure and $c_s$ is the (constant) sound speed.
We also assume that the background density is spatially uniform.
The background velocity then follows a Hubble-type law,
\be
{\bf u}_{\rm b}({\bf x},t)=
{\mathbb H}(t)\cdot{\bf x},
\label{grad}
\ee
where ${\mathbb H}(t)={\rm diag} ({\dot a}/a, {\dot b}/b, {\dot
c}/c)$,
and the equation of continuity implies
\be
\rho_{\rm b}(t)=\frac{\rho_0}{abc},
\ee
where $\rho_0=\rho_{\rm b}(t_0)$ is the initial value of the background
density. 
Let introduce a system of comoving coordinates
\be
{\bf x}^\prime=
{\mathbb S}^{-1}(t)\cdot {\bf x}.
\ee
in which the line element advected by the background flow appears stationary.
In these comoving coordinates the hydrodynamical equations become
\be
\frac{\partial \rho_1}{\partial t}+\tilde\nabla\cdot(\rho {\bf u}_1)=-{\rm tr}{\mathbb H}\,\rho_1,
\ee
and
\be
\frac{\partial {\bf u}_1}{\partial t} +({\bf u}_1\cdot \tilde\nabla){\bf u}_1 
+\frac{c_s^2}{\rho}\tilde\nabla \rho_1 +{\bf g}_1=-{\mathbb H}\cdot {\bf u}_1,
\label{ulin}
\ee
where the spatial gradient $\tilde\nabla$ is
\be
\tilde\nabla =
\frac{1}{a}\frac{\partial}{\partial x^\prime}{\hat{\bf e}}_x+
\frac{1}{b}\frac{\partial}{\partial y^\prime}{\hat{\bf e}}_y+
\frac{1}{c}\frac{\partial}{\partial z^\prime}{\hat{\bf e}}_z = {\mathbb S}^{-1}\cdot \nabla^\prime,
\ee
and the gravitational field ${\bf g}_1$ satisfies the Poisson's equation
\be
\tilde\nabla \cdot{\bf g}_1=4\pi G\rho_1.
\label{pois}
\ee

\section{Linear evolution of perturbations}

The linearised equations are
\be
\frac{\partial \delta}{\partial t}+\tilde\nabla\cdot {\bf u}_1=0,
\label{s1}
\ee
and
\be
\frac{\partial {\bf u}_1}{\partial t}+{c_s^2}\tilde\nabla\delta +{\bf g}_1=-{\mathbb H}\cdot {\bf u}_1.
\label{s2}
\ee
where $\delta({\bf x}^\prime, t)=\rho_1({\bf x}^\prime, t)/\rho_{\rm b}(t)$ is the density contrast.
Taking the curl ($\tilde\nabla\times$) of eq.~(\ref{s2}) we obtain the evolution of the vorticity 
$\boldsymbol{\omega}_1 ={\tilde\nabla}\times {\bf u}_1$, 
\be
\frac{\partial{\boldsymbol\omega}_1}{\partial t}=-{\rm tr}{\mathbb H}\,{\boldsymbol\omega}_1+{\mathbb H}\cdot {\boldsymbol\omega}_1,
\label{vort}
\ee
showing that any initial vorticity is enhanced by the contraction of the cloud (${\mathbb H}<0$).
If the contraction is isotropic, eq.~(\ref{vort}) reduces to 
\be
\frac{\partial{\boldsymbol\omega}_1}{\partial t}=-2\frac{\dot a}{a}{\boldsymbol\omega}_1,
\label{vortisot}
\ee
showing the the vorticity increases as $a(t)^{-2}$, the reverse 
of the so-called ``natural decay'' of vorticity in the expanding 
Universe. A more general form of eq.~(\ref{vortisot}) for the mean vorticity
$\langle {\boldsymbol\omega}^2\rangle^{1/2}$ has been obtained by 
Olson \& Sachs~(1973) who included a nonlinear term representing the breakup 
of larger eddies into smaller ones in incompressible turbulence 
(a process that tends to increase the mean vorticity). While in the expanding 
Universe the latter process competes with 
the ``natural decay'' due to the overall expansion, in a contracting cloud it
always leads to an unbounded increase of the mean vorticity that blows up in a 
finite time, if the fluid has zero viscosity (Olson \& Sachs~1973).
These ideas have been expanded by Robertson \& Goldreich~(2012).

Taking the divergence ($\tilde\nabla \cdot$) on both sides of eq.~(\ref{s2})
and using eq.~(\ref{s1}), we obtain the evolutionary equation for compressible modes 
(that couple to gravity)
\be
\frac{\partial^2\delta}{\partial t^2}-c_s^2\tilde\nabla^2\delta
-4\pi G\rho_{\rm b}\delta
=2\tilde\nabla\cdot({\mathbb H}\cdot{\bf u}_1).
\label{delta}
\ee
Consider a single Fourier mode with amplitude
\be
\delta({\bf x}^\prime, t)= F(t) 
e^{i{\bf k}^\prime\cdot{\bf x}^\prime},
\ee
where ${\bf k}^\prime$ is the (constant) comoving wavevector, related to the 
proper (time-dependent) wavevector ${\bf k}(t)$ by
\be 
{\bf k}(t)= 
{\mathbb S}^{-1}(t){\bf k}^\prime.
\label{eqk}
\ee
Inserting this expression in eq.~(\ref{delta}), we obtain
\be
\frac{d^2F}{dt^2}+2 H\frac{dF}{dt}+c_s^2(k^2 -k_{\rm J}^2)F=0,
\label{eqf}
\ee
where $k=|{\bf k}|$,
\be
H=\frac{1}{k^2}\left(\frac{\dot a}{a}k_x^2+\frac{\dot b}{b} k_y^2+\frac{\dot c}{c} k_z^2\right)
=-\frac{1}{k}\frac{dk}{dt}
\label{defh}
\ee
is the contraction rate, and 
\be
k_{\rm J}(t)=\frac{\sqrt{4\pi G\rho_b(t)}}{c_s}\equiv \frac{k_{\rm J,0}}{\sqrt{abc}}
\ee
is the Jeans wavenumber. The effect of the contraction is to
decrease the wavelength $\lambda(t)=2\pi/k(t)$ and increase the
amplitude (second term on the LHS, with $H<0$) of perturbations.
If the compression rate in one direction is much larger than in
the other two, perturbations grow faster along that direction and
become asymptotically two-dimensional.  The wavevector of the
perturbations, fixed in comoving coordinates, in physical space
becomes progressively aligned with the direction of stronger
compression. 

Defining  the variable $\xi$ as
\be
\frac{d\xi}{dt}=c_s\frac{k^2(t)}{k_{\rm J,0}^2}
\ee
eq.~(\ref{eqf}) can be written in compact form as
\be
\frac{d^2 F}{d\xi^2}+\left(\frac{k_{\rm J,0}}{k}\right)^4 (k^2-k_{\rm J}^2)F=0,
\label{eqc}
\ee
which generalises the ``basic equation of fragmentation theory'' by Lynden-Bell~(1973).

\subsection{WKB approximation}

If $k >k_{\rm J}$ perturbations oscillate.
If the oscillation period is much smaller than the contraction time, the amplitude
of perturbations can be estimated by a WKB analysis (see e.g. Falle~1972). Assuming
\be
F(\xi)=f(\xi) e^{-i\phi(\xi)}
\label{defF}
\ee
with $\phi(\xi)$ oscillating on a timescale much smaller than variation of $f(\xi)$,
the dominant terms in eq.~(\ref{eqc}) can be eliminated by choosing
\be
\frac{d\phi}{d\xi} = \left(\frac{k_{\rm J,0}}{k}\right)^2 (k^2-k_{\rm J}^2)^{1/2},
\label{dphi}
\ee
whereas the next largest terms give the condition
\be
f\frac{d^2{\phi}}{d\xi^2}+2\frac{df}{d\xi}\frac{d\phi}{d\xi}=0,
\ee
which implies $f^2\, d\phi/d\xi=$~constant.
From eq.~(\ref{dphi}), one then obtains 
\be
f \propto 
\frac{k}{(k^2-k_{\rm J}^2)^{1/4}}.
\label{ampli}
\ee
For $k\gg k_{\rm J}$, eq.~(\ref{defF})
represents oscillations with instantaneous frequency 
\be
\omega(t)=\frac{d\varphi}{dt}\approx c_s k(t)
\ee
and amplitude increasing as $k(t)^{1/2}$. 
If the contraction is isotropic with scale factor $a(t)$
the period of the oscillations ($k^{-1}\propto a$)
becomes progressively smaller than the timescale
of the variation of the amplitude ($k^{1/2}\propto a^{-1/2}$), thus making the WKB approximation
valid at any time, if it is satisfied initially, whereas in cosmology
the WKB approximation is satisfied only at early times (see, e.g. Peebles~1980).
Similarly, a WKB analysis of eq.~(\ref{s1}) shows that $|{\bf
u}_1|=c_s \delta\propto a^{-1/2}$.  In this approximation,
compressible modes conserve the action density
\be
\frac{{\cal E}(t)}{\omega(t)} = {\rm const.},
\ee
where
\be
{\cal E}(t) = \frac{1}{2} c_s^2 \delta^2 +\frac{1}{2}|{\bf u}_1|^2=c_s^2\delta^2
\ee 
is the energy density of the perturbations (Bretherton \& Garrett~1969,
Dewar 1970). 

\subsection{Special case: free-fall on a point mass}

\begin{figure}
\includegraphics[width=\columnwidth]{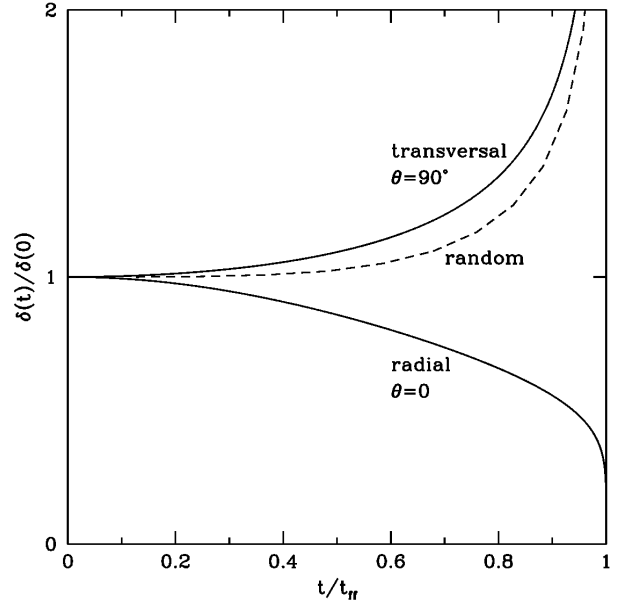}
\caption{Amplitude of density perturbations during spherical free-fall on a point mass, 
for perturbations with wavenumber radial  ($\theta=0$) and transversal ($\theta=90^\circ$). 
The dashed line shows the evolution of the amplitude for a random distribution of wavenumbers.}
\label{fig:nonisotropic}
\end{figure}

As an application, we evaluate the amplification of small-scale perturbations
($k \gg k_{\rm J}$)
during spherical free-fall on a star of mass $M_\star$. In
this case, the fluid element experiences a compression in the
transverse directions, say $x$, $y$, and a stretching in the radial
direction, say $z$,
\be
a(t)=b(t)=\cos^2\mu(t) \qquad c(t)=1+\frac{\sin^2\mu+3\mu\tan\mu}{2},
\ee 
where $\mu$ is a ``development angle'' ranging from 0 to $\pi/2$. The
development angle is related to time by 
\be
t=\frac{2}{\pi}(\mu+ \sin\mu\cos\mu)\,t_{\rm ff},
\label{time}
\ee
where $t_{\rm ff}=(\pi r_0^3/8GM_\star)^{1/2}$ is the free-fall time of a fluid element
initially at a distance $r_0$ from the star (see Appendix A2 for details).
From eq.~(\ref{ampli}), the amplitude of a density perturbation
propagating at an angle $\theta$ (fixed in the comoving frame) with
respect to the radial direction increases as
\be
\delta=\delta_0 \left(\frac{\sin^2\theta}{a^2}+\frac{\cos^2\theta}{c^2}\right)^{1/4},
\label{ampliff}
\ee
where the scale factors $a(t)$ and $c(t)$ are given by
eq.~(\ref{ffscales}).  Fig.~\ref{fig:nonisotropic} shows the evolution
of $\delta$ for perturbations with wavenumber radial  ($\theta=0$)
or transversal ($\theta=90^\circ$).
The dashed line shows the evolution of the amplitude for a random
distribution of wavenumbers
obtained averaging eq.~(\ref{ampliff}) over $\theta$. Since a fluid element
contracts in two directions while is stretched in the third,
the growth of perturbations is relatively slow: the average amplitude 
doubles only after 95\% of the free-fall collapse of the fluid element has 
been completed. An even smaller amplification can be expected in the accretion
flows on a filament, if the latter is approximated as an infinite cylinder, as 
contraction in this case only occurs in the direction perpendicular to the filament.

In general, anisotropic collapse and stretching of fluid elements in flows driven by 
the gravitational field of mass concentrations like 
stars or filaments dilute and retard the growth of any small-scale density/velocity perturbation 
initially present in the gas. In the next Section we turn to the collapse of a starless, 
self-gravitating cloud core where the amplification of perturbations is expected to
occur at the highest possible rate. 

\section{Application to homologous collapse}

The dynamics of starless cores appears to be characterised by a
overall slow contraction rather than free-fall collapse in their central 
parts (Keto et al.~2015, Campbell et al.~2016). Nevertheless, the simple model
of the homologous collapse of a uniform density, pressureless sphere
contains the essential characteristics of flows driven by the 
self-gravity of the gas (Whitworth \& Ward-Thompson~2001).
In addition, its simplicity allows one to derive analytically the time evolution
of the scale factors needed to apply the formalism of inverse Hubble
flows. In this model, the
acceleration of the background results from the self-gravity of the core,
\be
\nabla^\prime\cdot {\bf g}_{\rm b}= 4\pi G\rho_{\rm b},
\ee
where $\rho_{\rm b}=\rho_0/a^3$,
and the scale factor is 
\be
a(t)=\cos^2\mu(t),
\ee
where $\mu$ is again given by eq.~(\ref{time}) and 
$t_{\rm ff}=({3\pi/32G \rho_0})^{1/2}$ is the free-fall time
(see Appendix A1 for details).
In this case eq.~(\ref{eqf}) becomes
\be
\frac{d^2 F}{d\mu^2}-2\tan\mu\frac{dF}{d\mu}+6\left(\frac{k^{\prime 2}}{k_{{\rm J},0}^2}-\frac{1}{\cos^2\mu}\right)F=0.
\label{isowave_mu}
\ee

The evolution of density perturbations depends on their initial
spatial scale.  Consider for example the evolution of small-scale,
Jeans-stable perturbations with $k^\prime
\gg k_{{\rm J},0}$. In a static cloud, such fluctuations oscillate
as sound waves with frequency $\omega= c_s k$.  If  the cloud
contracts isotropically with scale factor $a$, the size of the
perturbation decreases as $a$ (its oscillation frequency increasing
accordingly as $a^{-1}$) but the Jeans length decreases faster, as
$c_s/\sqrt{\rho}\sim a^{3/2}$.  Thus, any small-scale perturbation
that is initially linearly stable and propagates as sound waves
becomes gravitationally unstable at some time $t_{\rm gr }(k^\prime)$,
that approaches $t_{\rm ff}$ as $k^\prime \rightarrow \infty$. From
eq.~(\ref{isowave_mu}), this time corresponds to $\cos^2\mu_{\rm
gr}=k_{{\rm J},0}^2/k^{\prime 2}$.  
In this limit, the solution of eq.~(\ref{isowave_mu}) is
\be
F(\mu)= \frac{1}{\cos\mu}[c_1\sin(p\mu)+c_2\cos(p\mu)],
\label{ff}
\ee
where $p=\sqrt{6}k^\prime/k_{{\rm J},0}\gg 1$, representing two
oscillating modes (with $F(0)=0$ and $dF/d\mu(0)=0$, respectively).
As anticipated by the WKB analysis, the amplitude of the perturbations
increase as $1/\cos\mu\propto a^{-1/2}$.

As the contraction of the cloud proceeds,
$a$ decreases, and eventually perturbations on scales progressively
small become gravitationally unstable. When
$k^\prime \ll k_{\rm J,0}$, eq.~(\ref{isowave_mu}) becomes
\be
\frac{d^2 F}{d\mu^2}-2\tan\mu\frac{dF}{d\mu}-\frac{6}{\cos^2\mu}F=0.
\label{isowave_mu_large}
\ee
General solution of eq.~(\ref{isowave_mu_large}) are well known from cosmology 
(e.g. Narlikar~2002). With the initial conditions appropriate for cloud collapse 
the solution is (Toal\'a et al.~2015),
\be
F(\mu)=c_1\frac{\tan\mu}{\cos^2\mu}+c_2\frac{2+\sin^2\mu+3\mu\tan\mu}{\cos^2\mu},
\label{gravinst}
\ee
and represents two growing modes (with $F(0)=0$ and $dF/d\mu(0)=0$,
respectively). Asymptotically for $t\rightarrow t_{\rm ff}$ both
modes grow like $a^{-3/2} \propto (1-t/t_{\rm ff})^{-1}$ (e.g.
Hunter~1962). Perturbations become gravitationally unstable at a
time $t_{\rm J}(\lambda^\prime)$ corresponding to the evolutionary
angle
\be
\cos\mu_{\rm J}=\frac{\lambda^\prime}{\lambda_{\rm J,0}}.
\ee
Thus, in a free-falling background, small-scale, Jeans-stable perturbations are
amplified as $a^{-1/2}$ until they become gravitationally
unstable and then grow as $a^{-3/2}$, the same rate 
characterising the evolution of the parent cloud, which is itself 
gravitationally unstable ($H\propto (1-t/t_{\rm ff})^{-1}$).

\subsection{Onset of nonlinearity}

As the amplitude of the perturbations increases, the linear approximation
breaks down at some point. Linear growth terminate when perturbations
become Jeans-unstable and collapse, or become nonlinear at a time $t_{\rm nl}$ 
and start to dissipate their energy by shocks. After $t_{\rm nl}$ the energy
dissipation is extremely rapid, as shocks efficiently convert wave energy 
into heat.

To evaluate $t_{\rm nl}$, consider for simplicity a one-dimensional flow.  Ignoring pressure
and gravity, the momentum equation (\ref{ulin}) reduces to 
\be
\label{EBM}
\frac{\partial u_1}{\partial t} + \frac{u_1}{a} \frac{\partial u_1}{\partial x^\prime}=-H u_1.
\ee
With the transformation $\varv_1=au_1$ and $d\tau=a^{-2}dt$, 
eq.~(\ref{EBM}) can be rewritten in the new variables $\varv_1(x^\prime,\tau)$ and $\tau$ 
as the standard (inviscid) Burgers' equation for the static case,
\be
\frac{\partial \varv_1}{\partial \tau} + \varv_1 \frac{\partial \varv_1}{\partial x^\prime}=0.
\label{bv}
\ee
The general (implicit) solution of eq.~(\ref{bv}) is easily obtained with the method of 
characteristics (see e.g. Jeffrey~2003):
a velocity perturbation with initial amplitude 
$\varv_1(x^\prime,0)=u_1(x^\prime,0)$ 
will steepen with time and form a shock at a time 
$\tau=- \{{\rm min}\,[du_1(x^\prime,0)/dx^\prime]\}^{-1} \equiv t_{{\rm nl},0}$,
where $t_{{\rm nl},0}$ is the time for reaching the nonlinear stage in the static case.
In the presence of contraction, nonlinearity therefore occurs at a time $t_{\rm nl}$ such that
\be
\int_0^{t_{\rm nl}}\frac{d\tau}{a^2} = t_{{\rm nl},0}.
\label{tnonlin}
\ee
Clearly, in the case of contraction $t_{\rm nl} < t_{{\rm nl},0}$,
i.e. perturbations enter the nonlinear phase earlier than in the static case. 

It is straightforward to evaluate $t_{\rm nl}$ in the case of the
homologous collapse of a pressureless sphere. A sinusoidal velocity perturbation 
with wavelength $\lambda^\prime=2\pi / k^\prime$ and initial amplitude $u_0$ 
becomes nonlinear and forms a shock at  $t=t_{\rm nl}$, corresponding to
the evolutionary angle $\mu_{\rm nl}$ obtained integrating
eq.~(\ref{tnonlin}),
\be
t_{{\rm nl},0}=\frac{4 t_{\rm ff}}{\pi}\int_0^{\mu_{\rm nl}} 
\frac{d\mu}{\cos^2\mu}=\tan\mu_{\rm nl}.
\ee
For $t_{{\rm nl},0}=1/ k^\prime u_0$, this implies
\be
\tan\mu_{\rm nl}=\frac{\lambda^\prime}{8 t_{\rm ff} u_0}.
\label{munonlin}
\ee

\begin{figure}
\includegraphics[width=\columnwidth]{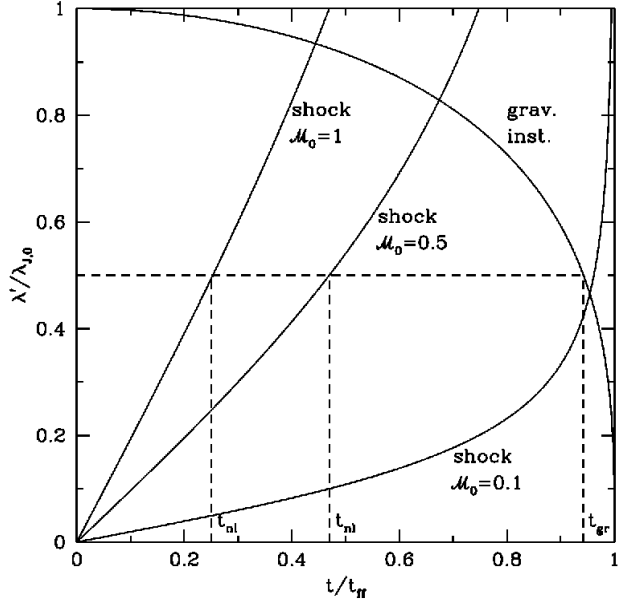}
\caption{During the collapse of a pressureless cloud, a linear
perturbation with comoving wavelength $\lambda^\prime$ becomes
nonlinear at a time $t_{\rm nl}$ ({\em curves labelled ``shock''})
and gravitationally unstable at a time $t_{\rm G}$ ({\em curve
labelled ``grav. inst.''}). Wavelength is in units of the Jeans
wavelength $\lambda_{\rm J,0}$ and time is in units of the cloud's
free-fall time $t_{\rm ff}$. The non-linear time is shown for three
values of the initial Mach number of the perturbations, ${\cal
M}_0=1$, ${\cal M}=0.5$ and ${\cal M}_0=0.1$.  The case of
$\lambda^\prime=0.5\,\lambda_{{\rm J},0}$ is shown as an example:
the nonlinear phase is reached before the gravitational instability,
unless the perturbations have Mach number below $\sim 0.1$.}
\label{fig:comparison}
\end{figure}

To compare $t_{\rm nl}$ and $t_{\rm gr}$, is convenient to parametrize
the amplitude of the perturbations as $u_0={\cal M}_0 c_s$, where
${\cal M}_0$ is the initial value of the rms Mach number, and rewrite 
eq.~(\ref{munonlin}) as
\be
\tan\mu_{\rm nl}=\left(\frac{1}{\sqrt{6}{\cal M}_0}\right)\frac{\lambda^\prime}{\lambda_{\rm J,0}}.
\ee
Fig.~\ref{fig:comparison} shows the time $t_{\rm nl}$ and $t_{\rm
G}$ at which initially Jeans-stable perturbations become nonlinear and gravitationally
unstable, respectively, for various values of the rms Mach number
representative of the level of non-thermal motions observed in 
dense cores, where typically ${\cal M}_0\approx 0.1$--1 (Myers~1983, Hacar \& Tafalla~2011).
Both $t_{\rm nl}$  and $t_{\rm gr}$ are 
bounded from above by $t_{\rm ff}$, but depend on the 
(comoving) scale of the perturbation $\lambda^\prime$ in opposite 
ways: $t_{\rm g}$ increases with decreasing $\lambda^\prime$, whereas 
$t_{\rm nl}$ becomes smaller. Thus, perturbations on sufficiently small scales
enter the nonlinear phase when they are still gravitationally stable.
For example, if $\lambda^\prime = 0.5
\lambda_{\rm J,0}$, the nonlinear stage and the formation of shocks
is reached at $t_{\rm nl}=0.26\, t_{\rm ff}$ and $t_{\rm nl}=0.47\,
t_{\rm ff}$ for ${\cal M}_0=1$ and $0.5$, respectively, well before
the perturbation becomes gravitationally unstable at $t_{\rm G}=0.94\,
t_{\rm ff}$.

\section{Contracting box model}

To follow the evolution of small-scale non-self gravitating
perturbations up to the nonlinear stage in a contracting background
we use \texttt{ECHO}, a {\it shock-capturing} code based on high-order
finite-difference methods for the evolution of classic and relativistic
magnetized plasmas (Londrillo \& Del~Zanna~2004, Del~Zanna et
al.~2007, Landi et al.~200.  The conservative nature of the
numerical method employed automatically preserves mass, momentum
and total energy density across any fluid discontinuity,  so that
the appropriate amount of dissipation is introduced at shocks to
capture their precise locations and to ensure the correct jump of
entropy.

The code works in general coordinate systems, including evolving
metrics, so that it can be easily adapted to expanding or contracting
numerical boxes, as done recently for a comprehensive study of the
nonlinear evolution of Alfv\'en waves in the expanding solar wind
plasma in the case of a uniform background speed (Del~Zanna et
al.~2015).  Here the method has been extended to any spatial metric
of the kind
\be
g_{ij} = \mathrm{diag}[a^2(t), b^2(t), c^2(t)],
\ee
leading to appropriate source terms in the (conservative) hydrodynamics
equations containing time derivatives of the above metric functions.

\begin{figure}
\includegraphics[width=\columnwidth]{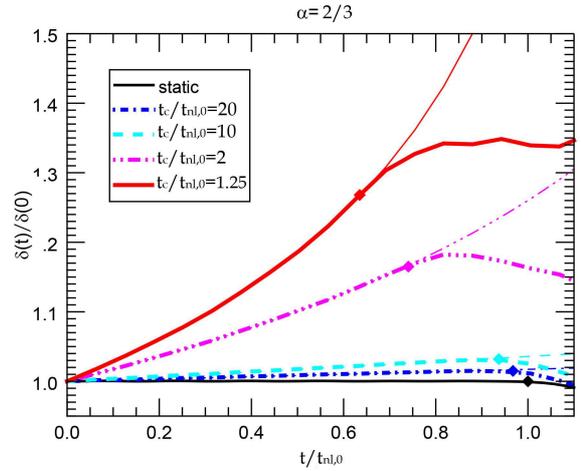} \caption{Time
evolution of the rms amplitude of density perturbations in a isothermal gas 
contracting isotropically 
with the scale factor of eq.~(\ref{scale}) with $\alpha=2/3$ (case of dynamical
contraction) for various values of $t_{\rm c}$ ({\em thick lines}). Ttime is in units
of $t_{{\rm nl},0}$, the nonlinear time for the static model.  A
{\em diamond} marks the position of $t_{\rm nl}$ given by eq.~(\ref{tnl}),
and {\em thin lines} show the evolution of the amplitude in the WKB approximation. 
For $t>t_{\rm nl}$ shocks are formed and the waves dissipate. }
\label{fig:case23}
\end{figure}

\begin{figure}
\includegraphics[width=\columnwidth]{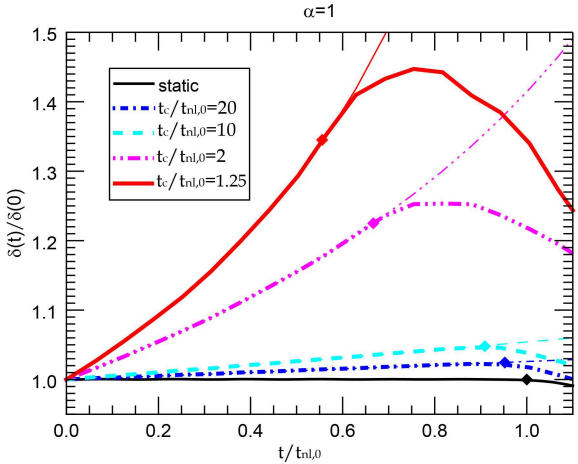}
\caption{Same as Fig.~\ref{fig:case23} for $\alpha=1$.}
\label{fig:case1}
\end{figure}

In the simulations, we have
assumed an isotropic contraction ($a=b=c$) from $t=0$ to $t=t_{\rm c}$ 
with scale factor
\be
a(t)=\left(1-\frac{t}{t_{\rm c}}\right)^\alpha.
\label{scale}
\ee 
The range of $\alpha$ of physical interest is $0<\alpha<2/3$, which
encompasses the case of a static background ($\alpha=0$), a
quasi-static collapse ($0<\alpha<2/3$), a dynamical collapse
($\alpha=2/3$). We also consider the linear case $\alpha=1$.
Fig.~\ref{fig:case23} and \ref{fig:case1} show the evolution of the density 
contrast, normalised to the initial value, for the $\alpha=2/3$
and $\alpha=1$, respectively, obtained with a Cartesian box of $512^3$
grid points. In the simulations, for a given $\alpha$ we fix the initial amplitude
of the perturbations to $\delta_0=0.1$ and take as reference time
the nonlinear time $t_{{\rm nl},0}$ for the static case. 
We then consider various contraction times, $t_{\rm c}/t_{{\rm nl},0}=20$, 10, 2, 1.25
and we follow the evolution of perturbations with time. 
We have run the same simulations with increasing spatial resolutions, 
observing that for $512^3$ grid points convergence was finally achieved. 
With the scale factor given by eq.~(\ref{scale}), the perturbations are 
expected to become nonlinear
at a time $t_{\rm nl}$ given by eq.~(\ref{tnonlin}),
\be
\frac{t_{\rm nl}}{t_{{\rm nl},0}}=\frac{t_{\rm c}}{t_{{\rm nl},0}}
\left\{1-\left[1-(1-2\alpha)\frac{t_{{\rm nl},0}}{t_{\rm c}}\right]^{1/(1-2\alpha)}\right\}.
\label{tnl}
\ee
Thus, the decay of perturbations 
corresponds to the physically correct amount of kinetic energy dissipation compatible with 
the $\gamma=1$ adiabatic index adopted in our simulations.

As shown by Fig.~\ref{fig:case23}, the evolution of perturbations depend on 
the time scale of the contraction of the box. The growth phase 
follows strictly the adiabatic approximation $\delta\propto a(t)^{-1/2}$
until nonlinear effects result in steepening of the waves and formation
of shocks. The analytical expression eq.~(\ref{tnl}) based on 
the one-dimensional inviscid Burgers' equation
predicts the end of the adiabatic phase and the beginning of the
dissipative phase, in good agreement with the numerical results.
The figure shows that $t_{\rm nl}$ is considerably reduced with
respect to the static case if the contraction occurs on a time
scale of the order of a few $t_{{\rm nl},0}$. If $t_{\rm c}\approx t_{\rm nl}$
the amplification due to contraction roughly balances the decay 
due to dissipation, maintaining an almost constant value of the
rms of density (and velocity) perturbations. 

\section{Discussion and Conclusions}

We have considered the evolution of small-scale hydrodynamic
perturbations in a contracting core, generalising the analytical approach
developed in cosmology for the expanding Universe 
to the case of arbitrary anisotropic inverse Hubble flows.
In general, gravity-driven anisotropic flows, as those characterising 
the accretion onto point-like or filament-like mass concentrations
are less effective than isotropic flows driven by the 
self-gravity of a cloud core in the amplification of density and 
velocity perturbations, due to the diluting effect of the stretching 
of fluid elements in the direction of the flow.

Even in the isotropic case, however, there are obvious differences with the expanding 
Universe, where both solenoidal and compressible modes decay because
of the expansion of the Universe, leaving only the gravitational
instability to promote the formation of substructures. In a contracting
cloud, solenoidal and compressible perturbations grow with time, 
the former faster than the latter
(as $a^{-2}$ or faster, in the isotropic case), promoting the the formation
of vortex structures preferentially aligned to the direction of
faster contraction. This process may create local enhancements
of angular momentum even in the absence of any bulk rotation of the
core, and core fragmentation may proceed from the break-up of
rotationally supported substructures, as found e.g. in the simulations
by Goodwin et al.~(2004). Conversely, the amplitude of compressible modes
in the WKB approximation grows 
as $a^{-1/2}$  in the isotropic case,
conserving the wave action density. These modes couple to the
gravitational field, and become gravitationally unstable as the
Jeans length progressively shrinks (as long as the collapse is isothermal), 
leading to fragmentation. 
However, the actual fate of compressible perturbations
depends in general on their wavelength and amplitude, and is controlled
by the relative values of several timescales: the time to reach the
nonlinear stage $t_{\rm nl}$, the time at which perturbations become
gravitationally unstable $t_{\rm gr}$, and the global collapse time
scale of the background.
The nonlinear time in a collapsing cloud,
bounded from above by the free-fall time $t_{\rm ff}$, is 
shorter than in a static background, and can be estimated  
analytically on the basis of a one-dimensional 
Burgers' equation modified by the contraction. For the typical 
amplitudes of velocity perturbations observed in cloud cores ($\delta u/c_s
\approx 0.5$--1), in general $t_{\rm nl} < t_{\rm gr}$, depending 
on their wavelength. Only perturbations on scales larger than $\sim 80$--90\% 
of the initial Jeans length (therefore of the order of the size of the core),
become Jeans-unstable in the linear phase. The others form shock
and start to dissipate their energy, establishing a competition with the 
``adiabatic heating'' resulting from contraction.

This scenario based on analytical results is supported by 
fully three-dimensional numerical calculations performed 
with the hydrodynamical code \texttt{ECHO}. These simulations confirm that
initially linear perturbations in a box contracting isotropically
increase their amplitude as predicted by the WKB analysis 
up to the time $t_{\rm nl}$.
As shown by the simulations, $t_{\rm nl}$ marks 
the onset of a phase of decay of the perturbations' amplitude due
to strong energy dissipation by shocks. If the 
rapid amplitude decay following $t_{\rm nl}$ 
suggested by the numerical calculations is representative of
the dissipation in cloud core, it is unlikely that small-scale
perturbations can survive to the point of becoming self-gravitating
and unstable. Under these circumstances, multiple fragmentation 
in cores is likely achieved by the growth and subsequent
breakup of solenoidal, rather than compressible, small-scale 
perturbations, either through the fragmentation of disk-like 
structures (Goodwin et al.~2004), or by the accumulation of 
mass at the boundary between nearby anti-parallel vortices
(Clarke et al.~2017). However, it should be kept in mind that 
vortices are easily disrupted by any large-scale magnetic field 
due to magnetic stresses associated to current sheets at 
their edges (Frank et al.~1996, Palotti et al.~2008),
an aspect that should be addressed in future works.
From the observational side, it would be desirable to exploit the 
capabilities of submillimetre interferometers like ALMA to constrain the level of 
fragmentation in starless cores, performing both high-sensitivity dust continuum
emission and molecular line observations, to probe the density
structure and the vorticity field prior to the formation of a 
multiple stellar system or a cluster. 

\section*{Acknowledgements}
CT acknowledges a CINECA award under the ISCRA initiative (project ISCRA C-TURCOL HP10C3MBH, 
for the availability of high performance computing resources and support.
LDZ and SL acknowledge support from the PRIN-MIUR project prot. 2015L5EE2Y 
{\em Multi-scale simulations of high-energy astrophysical plasmas}. 
The authors thank Paola Caselli, Roland Grappin, Victor Montagud-Camps, Marco Velli 
and Malcolm Walmsley for stimulating discussions.


\appendix

\section{Collapse and accretion}

\subsection{Homologous collapse of a pressureless sphere}

Consider a cloud of mass $M$ and uniform density $\rho_0$ collapsing in free-fall. 
The mass $M(r_0)$ inside a radius $r_0$ is $M(r_0)=(4\pi/3)\rho_0 r_0^3$,
and the free-fall time
is $t_{\rm ff}=(3\pi/32 G\rho_0)^{1/2}$, independent on the initial radius $r_0$. 
The radius of each shell as function of time is 
given by the parametric solution
\be
r=r_0\cos^2\mu, \qquad t=t_0(\mu+ \sin\mu\cos\mu),
\label{param}
\ee 
where $t_0=2 t_{\rm ff}/\pi$ and $\mu$ is a parameter (``development angle'') running from $0$ to $\pi/2$.  
Thus, all shells reach the origin at the same time, 
the density increasing uniformly as $\rho=\rho_0/\cos^6\mu$. Thus, in this case, 
\be
\delta r=\delta r_0 \cos^2\mu, 
\ee
and the contraction is isotropic with scale factors
\be
a(t)=w(t)=\cos^2\mu(t).
\ee
In particular,
\be
H(t)=\frac{\dot a}{a}=-\frac{\tan\mu}{t_0\cos^2\mu},
\ee
where $t_0=2t_{\rm ff}/\pi$. 

\subsection{Free-fall accretion flow}

Consider a fluid element free-falling on a star of mass $M_\star$. Let $r_0$ the position of 
the fluid element at time $t=0$ and $u=0$ its initial velocity. Then its velocity at a radius $r$ is 
\be
u(r)=-u_{\rm ff} \left(\frac{r_0}{r}-1\right)^{1/2},
\ee 
where $u_{\rm ff}=(2GM_\star/r_0)^{1/2}$.  Consider now a shell that at
time $t=0$ ($\mu=0$) has an outer radius $r_0$ and an inner radius
$r_0-\delta r_0$, where $\delta r_0 \ll r_0$. If the outer radius
of the shell reaches the star in a time $t_{\rm ff}=(\pi^2
r_0^3/8GM_\star)^{1/2}$, the inner radius reaches the star at $t_{\rm
ff}-\delta t$, where $\delta t/t_{\rm ff}\approx -3/2(\delta
r_0/r_0)$.  At any time $t(\mu)$, the parameter of the inner side
is $\mu+\delta\mu$, where
\be
r_0\, \delta\mu=\frac{3}{2}\left(\frac{\mu+\sin\mu\cos\mu}{1+\cos 2\mu}\right)\delta r_0.
\ee
Thus, while the outer radius of the shell is $r_0\cos^2\mu$, the
inner radius is $(r_0-\delta r_0)\cos^2(\mu+\delta\mu)$, and the
thickness of the shell during collapse is
\be
\delta r = \left(1+\frac{\sin^2\mu +3\mu \tan\mu}{2}\right)\delta r_0.
\ee
The radial scale factor $c(t)=\delta r(t)/\delta r_0$ is then 
\be
c(t)=1+\frac{\sin^2\mu +3\mu \tan\mu}{2}.
\ee
The rate of radial stretching is, after some algebra,
\be
\frac{\dot c}{c}=
\frac{3\mu + (3+2\cos^2\mu)\sin\mu\cos\mu}{t_0\cos^3\mu [\cos\mu(4+\sin^2\mu)+\sin\mu(6\mu+\sin\mu\cos\mu)]}.
\ee
On the other hand, the scale factor in the transversal direction is, as before, $a(t)=b(t)=\cos^2\mu(t)$.
Thus, a fluid element in free-fall on a star is stretched in the longitudinal direction and compressed in the transverse direction.
When $t\rightarrow t_{\rm ff}$, a first-order expansion gives
\be
a(t)=b(t) \rightarrow \left(1-\frac{t}{t_{\rm ff}}\right)^{2/3},
\qquad
c(t)\rightarrow \left(1-\frac{t}{t_{\rm ff}}\right)^{-1/3}.
\label{ffscales}
\ee
and
\be
\frac{\dot a}{a}\rightarrow -\frac{2}{3t_{\rm ff}}\left(1-\frac{t}{t_{\rm ff}}\right)^{-1}, \qquad
\frac{\dot c}{c}\rightarrow \frac{1}{3t_{\rm ff}}\left(1-\frac{t}{t_{\rm ff}}\right)^{-1}.
\ee
The rate of radial stretching is asymptotically  $1/2$ of the rate of transversal contraction and 
 is asymptotically equal to the radial velocity gradient $du/dr$.

\bsp    
\label{lastpage}
\end{document}